\documentclass[doublecol]{epl2}
\usepackage{bm}
\usepackage{amsmath}
\usepackage{amssymb}

\begin{document}

\title{Permeation through a lamellar stack of lipid mixtures}
\shorttitle{Permeation through a lamellar stack of lipid mixtures}

\author{Takuma Hoshino,\inst{1} Shigeyuki Komura\inst{1}
and David Andelman \inst{2}}
\shortauthor{T. Hoshino, S. Komura, and D. Andelman}

\institute{
\inst{1} Department of Chemistry, Graduate School of Science and Engineering,
Tokyo Metropolitan University, Tokyo 192-0397, Japan \\
\inst{2} Raymond and Beverly Sackler School of Physics and Astronomy,
Tel Aviv University, Ramat Aviv, Tel Aviv 69978, Israel\\
}


\abstract{
We study material transport and permeation through a lamellar stack of multi-component
lipid membranes by performing Monte Carlo simulations of a stacked two-dimensional Ising
model in presence of permeants.
In the model, permeants are transported through the stack via in-plane lipid clusters, which are
inter-connected in the vertical direction.
These clusters are  formed transiently by concentration fluctuations of the lipid mixture, and the
permeation process is affected especially close to the critical temperature of the binary mixture.
We show that the permeation rate decays exponentially as function of temperature
and permeant lateral size, whereas the dependency on the characteristic waiting time obeys
a stretched exponential function.
The material transport through such lipid clusters can be significantly affected around physiological
temperatures.
}

\maketitle

\section{Introduction}

Molecular transport and permeation in living systems are topics of growing interest in cosmetic
and drug delivery applications~\cite{G. Cevc}.
Such processes, characterized by selective transport of small ions and macromolecules
(biopolymers and proteins), often occur through stacks of biomembranes.
Examples of multi-lamellar structures within the cell are the Golgi apparatus
and mitochondria~\cite{AlbertsBook}.
In thylakoid membranes of chloroplasts, the ordered array of one of the photosystems
in a stack of membranes is responsible for photosynthetic functions such as energy
transfer and electron transport~\cite{Iwai2016}.
Moreover, it is known that different photosystems are heterogeneously distributed
between the stacked and unstacked regions of thylakoid membranes~\cite{Borodich2003}.

On a much larger tissue scale, an important system is the \textit{stratum corneum}, which
constitutes the outermost layer of human skin~\cite{Hadgraft2011}.
It is known that the stratum corneum is composed of corneocytes and intercellular lipids that form
lamellar structures~\cite{Hatta2011}.
Although stratum corneum lipids from a gel phase with limited mobility at physiological temperatures,
the lipid tail-tail interface of each bilayer is in a liquid-like disordered state~\cite{Das14}.
Cholesterol molecules incorporated in this liquid-like region can diffuse both translationally
and rotationally, which allows high overall cholesterol mobility.
Hence such inter-leaflet disordered regions can be regarded as heterogeneous fluid sheets
forming a multi-layered stack.

Using artificial stacks of multi-component lipid bilayers, Tayebi \textit{et al.}~\cite{Tayebi12}
reported that in-plane phase separation of lipid domains leads to an inter-layer columnar
ordering between the domains.
Such a strong vertical correlation between domains residing on adjacent membranes can lead
to a material transport, mediated through channel proteins that are preferentially
incorporated into these lipid domains.
However, it should be equally noted that a macroscopic phase separation usually does not occur in
biomembranes at physiological conditions, while thermal fluctuations of local concentrations are
always present in such multi-component
membranes~\cite{Smith08,Veatch08,Smith12,Ramachandran11,SK_DA_Review}.
The correlation length characterizing concentration fluctuations increases
as one approaches the critical temperature of the lipid mixture.
For example, it was shown that sub-micrometer  (about 50\,nm) concentration fluctuations
can take place in membranes at temperatures about 2--8$^\circ$C above their critical temperature~\cite{Smith08,Veatch08}.

In this Letter, we investigate the permeation process through a stack of two-component
(saturated and unsaturated) lipid membranes that exhibit strong concentration fluctuations
in the thermodynamically stable one-phase region (above the critical temperature).
We regard the multi-layered binary membranes as a stack of two-dimensional (2D) Ising
model, and consider their inter-layer correlations~\cite{Hoshino15}.
The permeable molecules residing on one membrane can be transferred, within our model,
to the adjacent membrane through transiently connected clusters bridging the two
neighboring membranes.
Using Monte Carlo simulations, we investigate in detail the dependency of the permeation
rate on temperature, permeant size, and the characteristic waiting time.

An important effect that is apparent close to the critical point is an enhanced permeability
through stacks of membranes because the life-time of transiently connected clusters along
the stack becomes longer due to critical slowing down.
We demonstrate that the permeation rate increases exponentially as the temperature approaches the
critical temperature from above, and conclude that this rate is related to the in-plane concentration
correlations.
Furthermore, the permeation rate is shown to decay exponentially as a function of
permeant lateral size.
We predict that concentration fluctuations in physiological conditions can play an important
role for efficient material transport through multi-component membranes.

\section{Model}

\begin{figure}[tbh]
\centering
\includegraphics[scale=0.3]{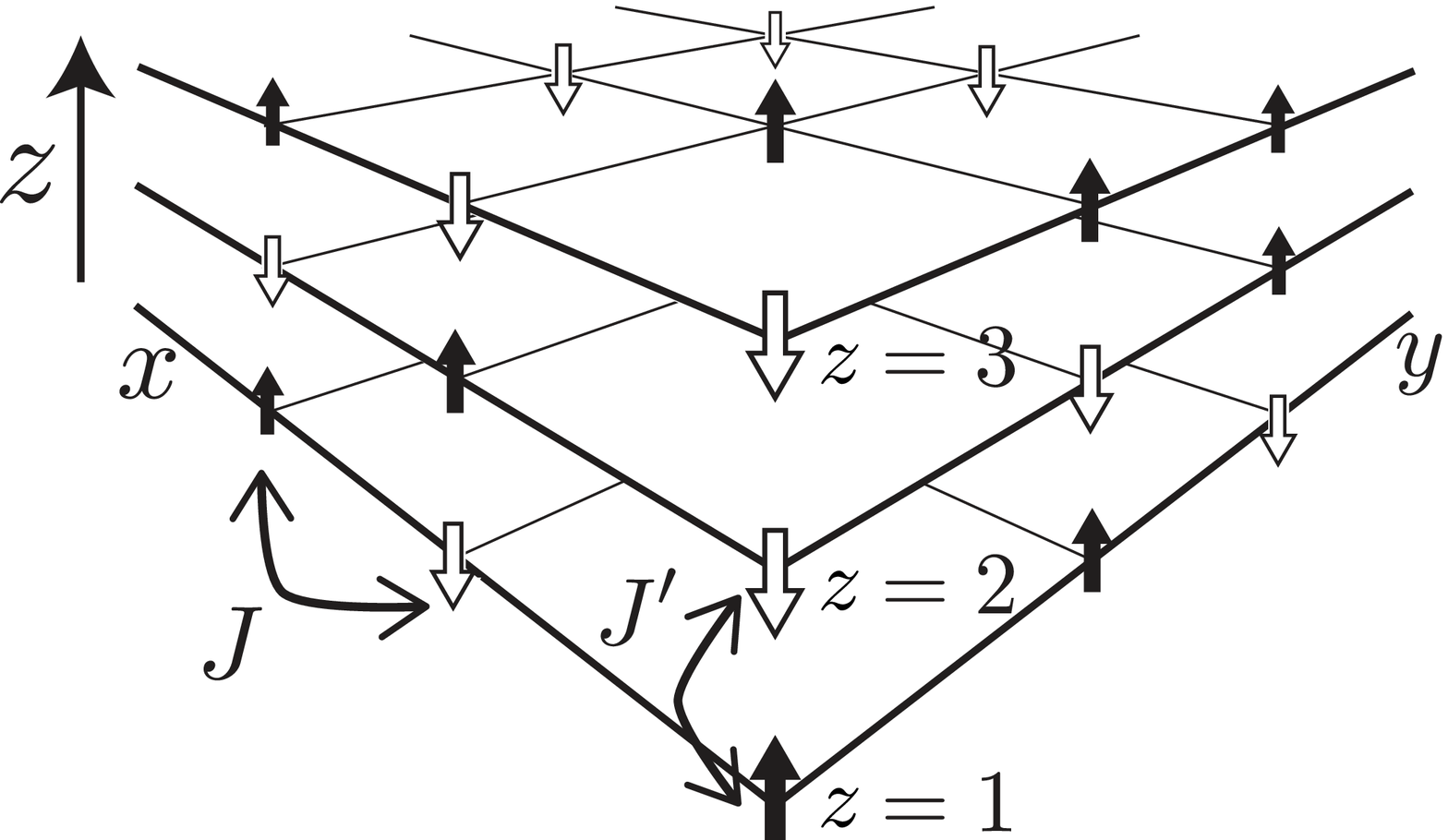}
\caption{\textsf{
The stacked 2D Ising model used to model a stack of membranes when the inner structure
of the membranes is ignored~\cite{Hoshino15}.
Each membrane is composed of a mixture of saturated lipids (S) and unsaturated lipids (U).
The S and U lipids correspond to spin up (black) and spin down (white), respectively.
The coupling parameter between nearest-neighbor spins (lipids) in the same layer is $J$,
and the coupling parameter between spins (lipids) belonging to two nearest-neighboring layers is $J'$.}
}
\label{fig1}
\end{figure}

As shown in fig.~\ref{fig1}, we consider a stack of lipid membranes composed
of a mixture of saturated (S) and unsaturated (U) lipids, modeled via the \textit{stacked 2D Ising model}.
Each lipid bilayer has a finite thickness and can be mapped into a 2D Ising model with conserved
magnetization corresponding to the average S/U lipid composition.
The Hamiltonian of the membrane stack is~\cite{Hoshino15}
\begin{align}
H = & -J \sum_{i, \langle \boldsymbol{\rho}, \boldsymbol{\rho}' \rangle}
S_{i,\boldsymbol{\rho}} S_{i,\boldsymbol{\rho}'}
-J' \sum_{i, \boldsymbol{\rho}}
S_{i,\boldsymbol{\rho}} S_{i+1,\boldsymbol{\rho}},
\label{hamiltonian}
\end{align}
where the spin variable $S_{i,\boldsymbol{\rho}}=\pm 1$ is located at in-plane position
$\boldsymbol{\rho}=(x,y)$ of the $i$-th layer along the $z$-direction, and corresponds to a
lattice site occupied either by an S  or U lipid, respectively.
Furthermore, $J$ is the in-plane coupling parameter between nearest-neighbor spins (lipids)
in the same layer,
{while $J'$ is the coupling parameter between spins (lipids) belonging to two
nearest-neighboring layers, originating primarily from direct van der Waals attractive
interactions~\cite{Israelachivili}.}
Throughout this Letter, we make use of the inter-layer coupling strength, defined as the
dimensionless ratio $\lambda \equiv J'/J$.

\begin{figure}[tbh]
\centering
\includegraphics[scale=0.3]{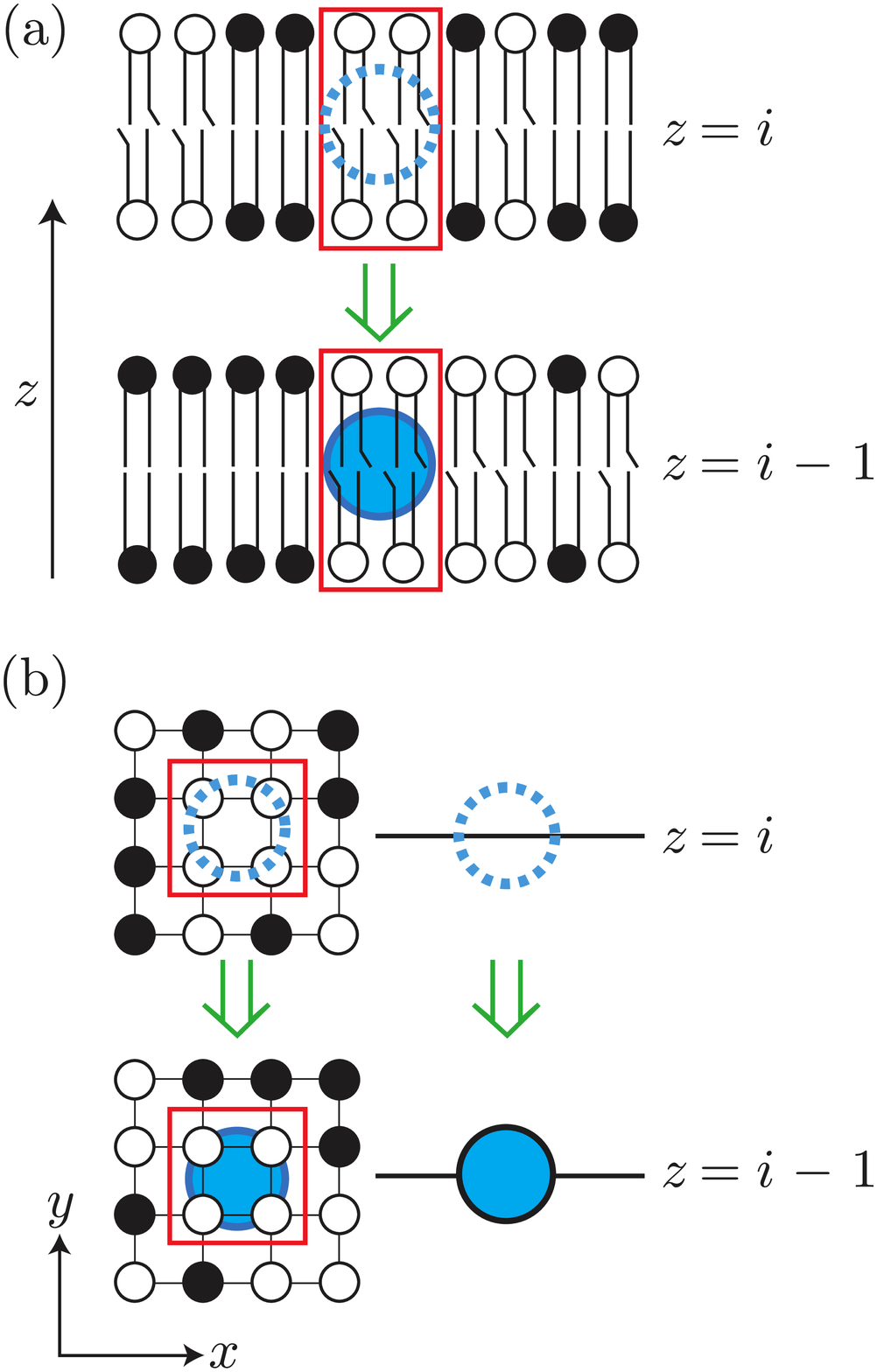}
\caption{\textsf{
(a) Schematic illustration along the $z$-axis of a stack of lipid membranes in the presence of permeant molecules.
Black and white lipids correspond to S and U lipids, respectively.
(b) A schematic representation of the permeation model (top view).
A permeant molecule is allowed to move from the $z=i$ layer to the $z=i-1$ layer below it,
if the following conditions are satisfied:
(i) A permeant is initially incorporated  in a lipid cluster at the top layer, $z=L_z$.
(ii) The lipid clusters should be connected in the $z$-direction across neighboring layers.
(iii) If the formation of clusters in the $z$-direction persists over waiting time of $t_w$ MCS,
the permeant is moved from the $z=i$ layer to the $z=i-1$ layer below it.
(iv) The procedure is repeated, in a unidirectional way, till all permeants reach the bottom layer of the stack, $z=1$.}
}
\label{fig2}
\end{figure}

To simulate the behavior of permeants in a stack of lipid membranes, Monte Carlo simulations are
performed for the equivalent Ising spin model, eq.~(\ref{hamiltonian}).
The simulations are done on a finite 3D lattice  of size $L\times L\times L_z$,
with periodic boundary conditions in all three spatial directions.
The spin configurations are updated using Kawasaki exchange dynamics, in order to conserve
the S/U lipid composition (the magnetization for the Ising model) in each layer~\cite{Hoshino15}.
Spin exchanges are only allowed for spin pairs that belong to the same layer,
because the time-scale of lateral diffusion is
much faster than that of out-of-plane lipid exchange between neighboring layers.
The probability of spin exchange is determined by the standard Metropolis algorithm.
{The important difference between the present stacked 2D Ising model and the
ordinary anisotropic 3D Ising model~\cite{Lee02} is that in the former the magnetization
(corresponding to the lipid composition) in each layer is conserved.
The thermodynamical properties such as the $\lambda$-dependent critical temperature of the
stacked 2D Ising model were studied in detail in Ref.~\cite{Hoshino15}.}

Next, we address the permeation process in the model.
In addition to the spin variable that accounts for the S/U mixture, we introduce another
variable to model the permeant molecules.
We assume that each permeant has a square shape of $m=n^2$ lattice sites and
lies flat within one layer.
In other words, its thickness is comparable to that of the lipid bilayer and will not be further
considered.

As schematically shown in fig.~\ref{fig2}, we require the following conditions for a permeant
to be transferred to the neighboring layer below.
(i) Initially, all permeant molecules are incorporated in a lipid cluster composed of S or U lipids,
located in the top layer, $z=L_z$.
(ii) The Monte Carlo simulation is run continuously till the S (or U) lipid cluster in the upper layer
lies above another cluster of the same S (or U) type in the adjacent layer below.
(iii) The transient overlap between the two clusters along the $z$-direction should  persist over
$t_w$ Monte Carlo steps (MCS).
Once the conditions (i)-(iii) are satisfied, the permeant molecule is moved down to the adjacent layer
below, while preserving the permeant lateral position.
(iv) The procedure is repeated, in a unidirectional way (no backward movement is allowed), till
all permeants reach the bottom layer of the stack, $z=1$.

The characteristic \textit{waiting time},  $t_w$
in units of MCS, introduced above is closely related
to the diffusivity or chemical affinity of the permeant (see Discussion below).
If the cluster connectivity is destroyed within the waiting time $t_w$, either by lateral
motion or disassembly of the lipid cluster, the permeant is forced to stay in the same layer.
Furthermore, the permeant lateral position is fixed because only the relative motion between
permeants and the lipid cluster is important.
{Note that we do not take into account any interaction acting between
permeant molecules, which is justified in the dilute limit of permeants.}
We also assume that the above permeation process is unidirectional (always permeates from
top to bottom), because protein machines such as ion channels are responsible for directed
transport of specific materials in biological membranes.
{In experiments, such transport processes may be realized by applying external forces
(\textit{e.g.}, electric field or concentration gradient) to permeant molecules.}

One MCS contains  $L\times L\times L_z$ spin updates, and the first 5,000 MCS are discarded
in order to reach thermal equilibrium.
After equilibration, we allow the permeants to penetrate through the layers according to the
algorithm introduced above.
The same permeation procedure is repeated by running several Monte Carlo runs in order to
improve the statistics.
The average S/U composition is taken to be at its critical value, \textit{i.e.}, 1:1 mixture of
S and U lipids.
The inter-layer coupling strength is fixed to $\lambda=J'/J= 0.1$, resulting in a critical
temperature $k_{\rm B}T_{\rm c}/J \approx 2.85$~\cite{Hoshino15}, where
$k_{\rm B}$ is the Boltzmann constant.
We emphasize that all Monte Carlo simulations are conducted for temperatures larger
than $T_{\rm c}$, namely, $T>T_{\rm c}(\lambda)$.

\section{Results}

\begin{figure}[tb]
\centering
\includegraphics[scale=0.3]{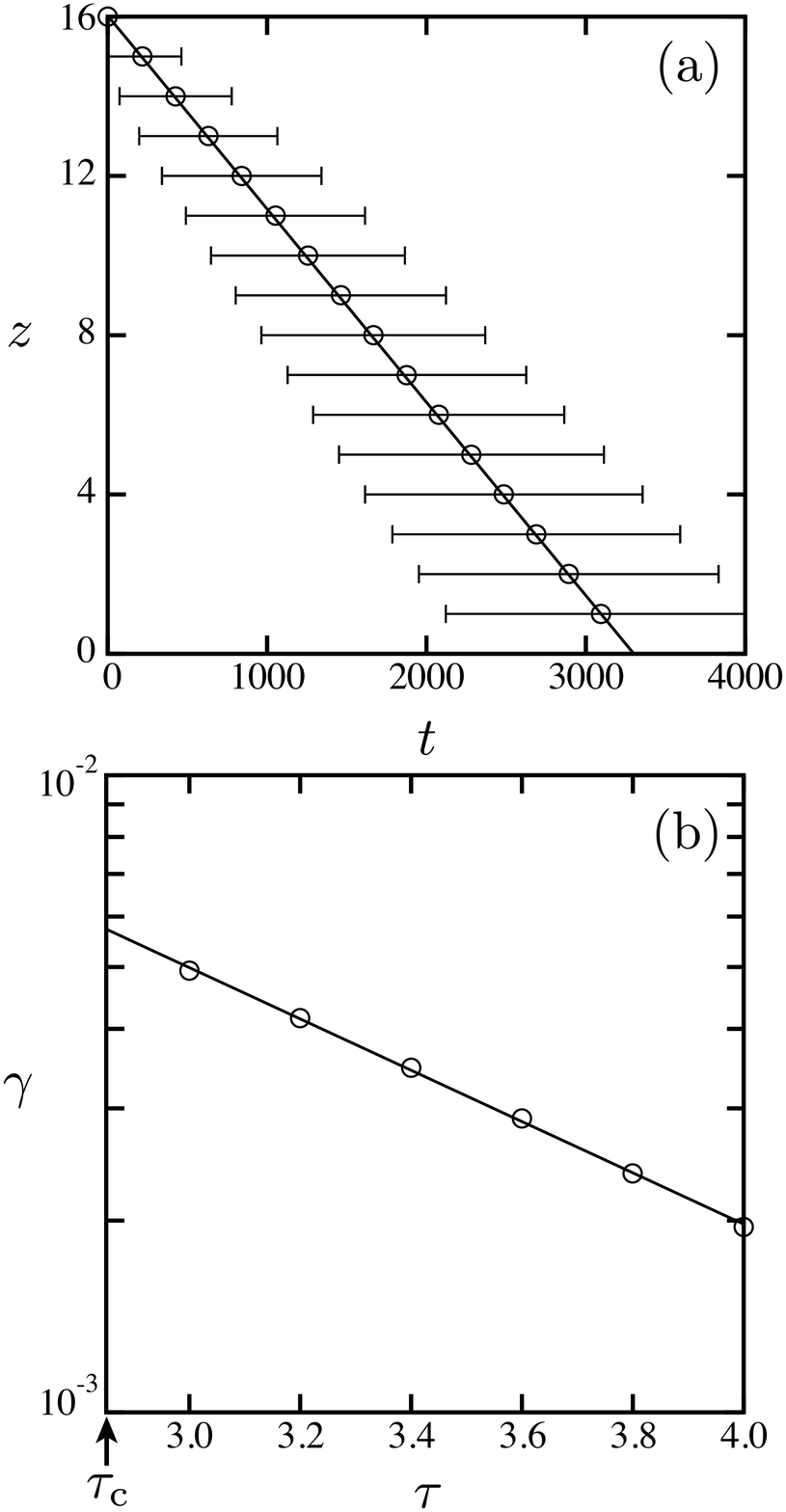}
\caption{\textsf{
(a) Average $z$-position of permeants as a function of time $t$ measured in units
of MCS after the initial 5,000 MCS are discarded for equilibration purposes.
The parameter values are $\tau=k_{\rm B}T/J=3.0$, $m=4$, $t_w=5$
{(in units of MCS)} and $\lambda=J'/J=0.1$.
The system size is $L\times L\times L_z=48\times 48\times 16$.
The data can be fitted by a linear relation as given by eq.~(\ref{z_vs_t}), with a
permeation rate $\gamma=4.9\times10^{-3}$.
The bars indicate the standard deviations obtained by averaging over 20 independent runs.
(b) The permeation rate $\gamma$ as function of the scaled temperature $\tau$ for $m=4$ and $t_w=5$.
The exponential decay of $\gamma$, as in eq.~(\ref{v_eT}), is obtained with
fitting coefficients, $A=0.9$ and $B=-2.5$.
Note that the critical temperature value for
$\lambda=0.1$ is $\tau_{\rm c}=2.85$ (shown by the arrow). This value is lower than
the temperatures used in the runs, which are always within the one-phase region.
The fitting error is within the size of the symbol.}
}
\label{fig3}
\end{figure}

In fig.~\ref{fig3}(a), we plot the average $z$-position of permeants (measured in units of the layer number)
as function of time $t$ (measured in units of MCS after discarding the initial 5,000 MCS) for scaled
temperature, $\tau \equiv k_{\rm B}T/J=3.0>\tau_{\rm c}=2.85$, permeant size $m=4$,
and waiting time $t_w=5$.
The obtained result can be fitted by a linear relation between  position $z$ and
time $t$
\begin{equation}
z=-\gamma t+L_z,
\label{z_vs_t}
\end{equation}
where the slope $\gamma$ corresponds to the \textit{permeation rate} -- to be
distinguished from the \textit{permeability coefficient} introduced earlier
in the literatures~\cite{Scheuplein1971,Potts92}.
Here the least-square fitting yields $\gamma=4.9\times10^{-3}$,
meaning that it takes about $\gamma^{-1} \approx 200$~MCS for a permeant
(of size $m=4$) to move down from the $i$-th layer to the adjacent $(i-1)$-th layer.
{According to the proposed Monte Carlo procedure as described above,
the permeant molecules are moved down only to the adjacent layer below within a single MCS.
Hence the above permeation rate should be always smaller than the inverse of the waiting time
$t_w$, \textit{i.e.}, $\gamma < 1/t_w$.}
The linear dependency in time shown in eq.~(\ref{z_vs_t}) is a consequence of the unidirectional
permeation process that is an important assumption within our model.
Moreover, $\gamma$ is roughly a constant during the permeation process,
since the average time needed to move from the $i$-th layer to the $(i-1)$-th one is the
same for any $i$.
On the other hand, $\gamma$ depends on the reduced temperature $\tau$,
the permeant size $m$, and the waiting time $t_w$, \textit{i.e.}, $\gamma=\gamma(\tau, m, t_w)$.
In the following, we shall examine these dependencies in more detail.

We first focus on the temperature dependency, and plot the permeation rate $\gamma$ as a
function of $\tau>\tau_{\rm c}$ on a semi-log plot in fig.~\ref{fig3}(b) for $m=4$ and $t_w=5$.
Note that the scaled critical temperature,
$\tau_{\rm c}=2.85$, is below the range of $\tau$ we examined.
As can be seen in fig.~\ref{fig3}(b), the obtained temperature dependency can be well-fitted
by an exponential form
\begin{equation}
\gamma(\tau, m, t_w)=\exp[-A(m, t_w)\tau +B(m, t_w)],
\label{v_eT}
\end{equation}
where the fitting coefficients $A$ and $B$ are functions of $m$ and $t_w$.
Notice that $A$ is mostly positive, indicating that the permeation rate $\gamma$
decreases as the scaled temperature, $\tau>\tau_{\rm c}$ increases.
In other words, the permeation rate increases exponentially as the temperature approaches
the critical temperature from above.
This is an important result and holds quite generally in our simulation.
Such a temperature dependency of $\gamma$ can later be explained in terms of the
correlation length which characterizes the lipid cluster size.

The above results indicate that in stacked membranes, lipid concentration fluctuations  significantly
affect the permeability, especially close to the critical point, $\tau \gtrsim \tau_{\rm c}$.
As $\tau\to\tau_{\rm c}$, the correlation length substantially increases and even diverges.
{This leads to a large increase of the relaxation time of those clusters (critical slowing down),
which can easily exceed the waiting time, $t_w$.}
Since the clusters in adjacent layers are strongly correlated even for a small value of the
coupling parameter $\lambda=J'/J=0.1$~\cite{Hoshino15}, the permeation rate
depends strongly on temperature as in eq.~(\ref{v_eT}).

\begin{figure}[tb]
\centering
\includegraphics[scale=0.3]{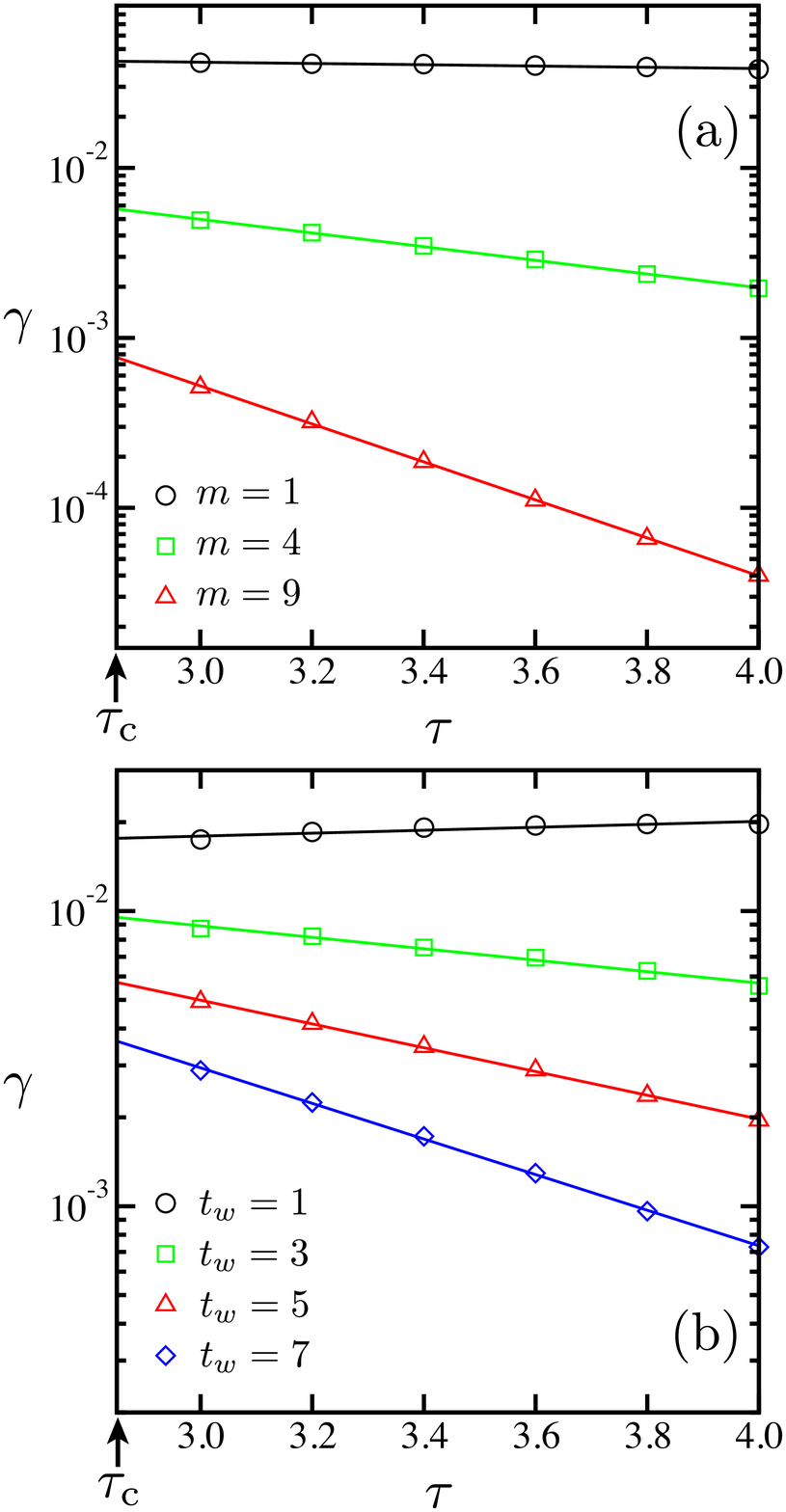}
\caption{\textsf{
(a) The permeation rate $\gamma$ as function of the scaled temperature $\tau$ for
different permeant size, $m=1, 4$ and $9$.
Other parameter values are $t_w=5$ and $\lambda=0.1$.
(b) The permeation rate $\gamma$ as function of the scaled temperature $\tau$
for different waiting times $t_w=1, 3, 5$ and $7$.
The other parameter values are $m=4$ and $\lambda=0.1$.
In both (a) and (b), $\gamma$ decays exponentially with $\tau$ except when
$m=1$ in (a) and $t_w=1$ in (b).
}}
\label{fig4}
\end{figure}

In fig.~\ref{fig4}(a), we plot  the permeation rate $\gamma$ as function of the
temperature $\tau$ for different permeant sizes $m=1, 4$\, and 9, while fixing $t_w=5$.
Although $\gamma$ hardly depends on $\tau$ for $m=1$,  it exhibits strong temperature
dependency for $m=4$ and $9$.
{Notice that $m=1$ corresponds to permeants that have the minimal size allowed
in the present simulation.}
On the other hand, for any fixed temperature above $\tau_{\rm c}$,  the decrease in $\gamma$
is found to be significant as the permeant size $m$ is increased.
This can be understood since it is more difficult for larger clusters to satisfy our above-mentioned
permeant transport conditions, (i) and (ii).
In fact, $\gamma$ decreases  nearly exponentially with $m$ when fixing the values of all other
parameters.
This result implies that the cross-section area of a permeant, $m=n^2$, is the important factor
in controlling the permeation because the area occupied by a permeant is kept constant while
it is transported to the neighboring layer.
Such a size dependency is in accordance with the experimentally observed exponential dependency
of the skin permeability coefficient on molecular volume~\cite{Potts92}.

In fig.~\ref{fig4}(b), we plot $\gamma$ as a function of the temperature $\tau$ for
different waiting times $t_w=1, 3, 5$\, and 7, while fixing $m=4$.
{We remark that an exponential decay is found except for $t_w=1$, which is the smallest
waiting time allowed in the model.
Although $\gamma$ is almost independent of $\tau$ for $t_w=1$ and $m=4$ due to
these small values, $\gamma$ generally decreases with $\tau$ for larger values of $m>4$
even for $t_w=1$ (not shown here).}
Furthermore, $\gamma$ is found to decrease as $t_w$ increases for any fixed temperature.
According to the permeation condition (iii), lipid clusters in two neighboring layers need
to overlap vertically over $t_w$ time steps in order to achieve the permeation.
This implies that the permeation rate $\gamma$ should be a decreasing function of $t_w$,
which is confirmed in our simulation.
However, as discussed below, the dependency of $\gamma$ on $t_w$ is not as simple as its
dependency on $\tau$ or $m$.

\begin{figure}[tb]
\centering
\includegraphics[scale=0.3]{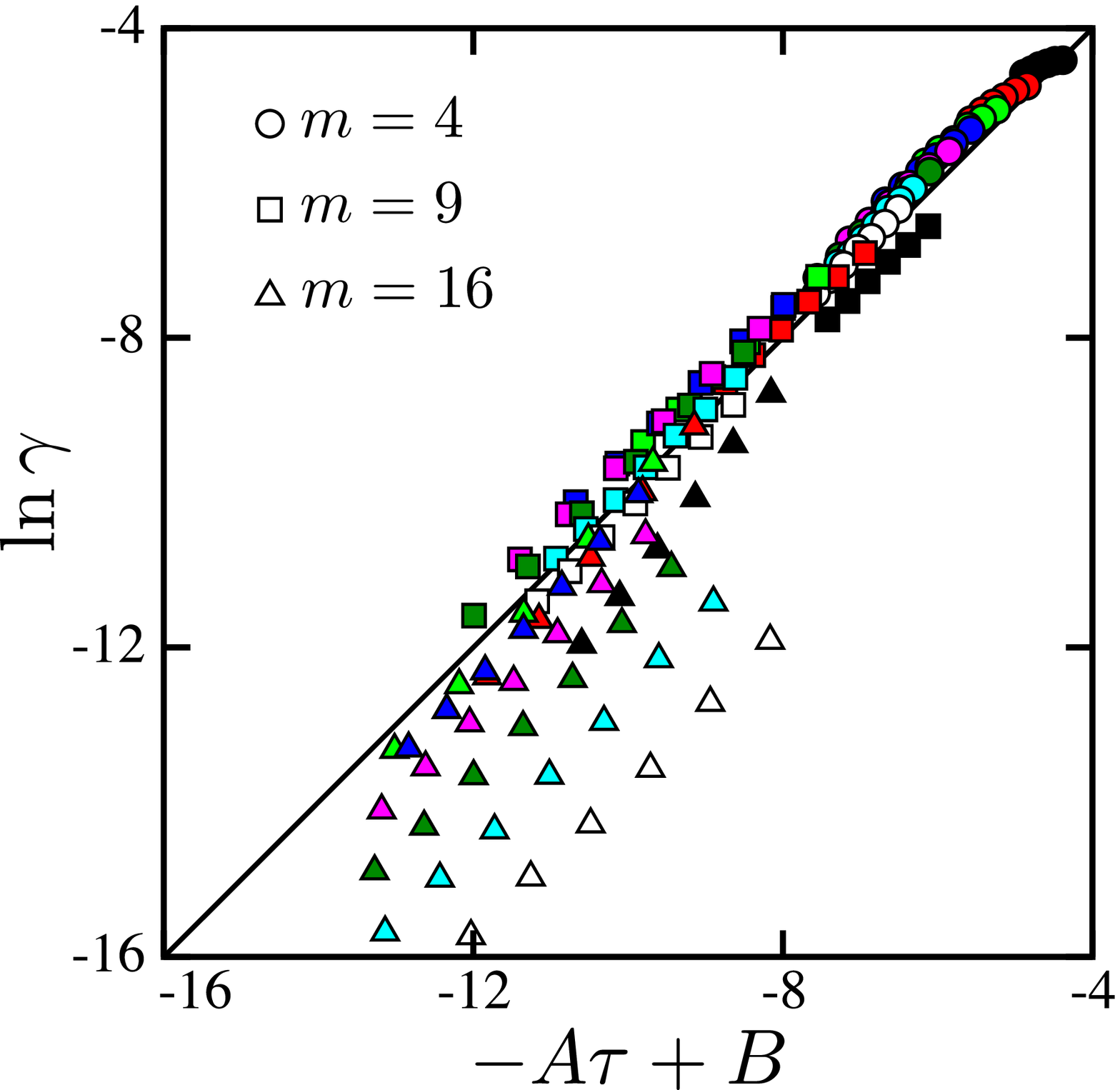}
\caption{\textsf{
Scaling plot of $\ln \gamma$ as a function of $-A\tau +B$ for $m=4, 9$ and $16$, and
$t_w=2, 3, \dots, 9$.
See eqs.~(\ref{v_eT}), (\ref{Eq.4A}), and (\ref{Eq.4B}).
Different symbols indicate different $m$ values (shown in the graph), while different colors
indicate different $t_w$ values;
$t_w= 2$ (black), $3$ (red), $4$ (light green), $5$ (blue), $6$ (magenta), $7$ (deep green),
$8$ (cyan), and $9$ (white).
}}
\label{fig5}
\end{figure}

To discuss more quantitatively the behavior of $\gamma(\tau, m, t_w)$ on $m$ and $t_w$,
we analyze the functions $A(m,t_w)$ and $B(m,t_w)$ of eq.~(\ref{v_eT}).
For sake of simplicity, we assume that both $A$ and $B$ obey the following
scaling forms:
\begin{align}
A(m,t_w) & \approx a_1 m^{\alpha_1}(t_w)^{\alpha_2} +a_2,
\label{Eq.4A} \\
B(m,t_w) & \approx b_1 m^{\beta_1}(t_w)^{\beta_2}+b_2,
\label{Eq.4B}
\end{align}
with four exponents $\alpha_1$, $\alpha_2$, $\beta_1$, and $\beta_2$ and four constants
$a_1$, $a_2$, $b_1$ and $b_2$.
The numerical fitting yields the following values:
$\alpha_1 \approx 1.03 \pm 0.095$,
$\alpha_2 \approx 0.74 \pm 0.067$, and
$\beta_1 \approx 1.61 \pm 0.16$,
$\beta_2 \approx1.37 \pm 0.16$.
Then, the leading combined dependency of $\gamma$ on the three parameters, $\tau, m$, and $t_w$ can be
written as $\gamma \sim \exp [- m (t_w)^{3/4} \tau]$, where only the $A$-part dependency is written explicitly.
This functional dependence on the three parameters is the main result of this Letter.

To check the validity of the above-proposed scaling form, we have plotted in fig.~\ref{fig5}
the quantity $\ln\gamma$ as a function of $-A\tau +B$ with $A$ and $B$ assumed to obey the scaling 
form as in eqs.~(\ref{Eq.4A}) and (\ref{Eq.4B}) for $m=4,\, 9$\, and 16, and $t_w=2, 3, \dots, 9$.
A good data collapse can be seen for smaller $m$ values, while there is a systematic deviation below
the fitting line for $m=16$ (triangles) as $t_w$ is increased.
Note that even in the latter case of $m=16$,
a linear relation between $\ln \gamma$ and $-A\tau +B$ is
maintained, indicating that the scaling form of $B$ in eq.~(\ref{Eq.4B})  becomes
inaccurate for larger $m$ and $t_w$.

\section{Discussion}

The permeation rate $\gamma$, as is obtained from our simulation, decays exponentially with
the permeant size $m$ and temperature $\tau$.
On the other hand, the dependency on the waiting time $t_w$ is described by a stretched
exponential with an exponent $\alpha_2 \approx 3/4$.
Our results are consistent with a stretched exponential relaxation associated with dynamic heterogeneity
that can be found in glassy systems.
In the trapping model~\cite{Phillips96}, a stretched exponential behavior with an exponent
$3/5$ was predicted for a dynamical correlation function in 3D.
Our obtained value $\alpha_2\approx 3/4$ is slightly larger, and the physical meaning of the precise
$\alpha_2$ value needs further clarifications.

We point out that the temperature dependency of the permeation rate in eq.~(\ref{v_eT}) is
analogous to that of the in-plane correlation function of an Ising spin system.
For the one-phase region ($T>T_{\rm c}$), the correlation function is given by
\begin{equation}
\langle S_{i,\boldsymbol{\rho}}S_{i,\boldsymbol{\rho}+\mathbf{r}}\rangle
\sim \exp(-r/\xi),
\label{corr_spin}
\end{equation}
where $r =\vert \mathbf{r} \vert$ and $\xi$ is the correlation length.
Close to the critical point, the correlation length as function of temperature scales as
$\xi \sim \vert \tau - \tau_{\rm c} \vert^{-\nu}$, where $\nu$ is the corresponding
critical exponent.
In our previous work~\cite{Hoshino15}, we have shown that the critical behavior of the
stacked Ising model can be described by a 2D Ising model even in the
presence of a coupling between adjacent layers.
Hence, the exponent $\nu$ should take the exact 2D value of $\nu=1$~\cite{Onsager44},
as has been also confirmed in our supplementary simulation (not shown here).

Therefore, the in-plane correlation length is inversely proportional to the reduced temperature,
\textit{i.e.}, $\xi \sim 1/(\tau - \tau_{\rm c})$, and the correlation function in
eq.~(\ref{corr_spin}) decays exponentially with temperature.
It is reasonable to expect that the temperature dependency of the permeation rate is essentially
the same as that of the correlation function, because only the permeants incorporated
in lipid clusters can be transferred to the adjacent membrane.

{Another finding of Ref.~\cite{Hoshino15} is that the correlations in the $z$-direction
are very strong because of the cooperative behavior of domains in different layers.
This feature arises because of the constraint that the lipid composition in each layer is strictly conserved.
As a result,  below the critical temperature ($\tau<\tau_{\rm c}$), the system forms a continuous
columnar structure for any finite interaction $\lambda>0$ across adjacent layers.

Although so far we have discussed concentration fluctuations above the critical temperature
($\tau \gtrsim \tau_{\rm c}$), we consider that the correlations in the $z$-direction are always
strong enough, especially close to the critical temperature.
Hence, we expect that the in-plane lateral correlation is a governing factor for the temperature dependence of 
the permeation rate $\gamma$.
In the present work, we have studied only the case of $\lambda=0.1$.
We think that this relatively small coupling parameter is sufficient to have strong correlations
in the $z$-direction.
To understand the general behavior, however, it is necessary to investigate the permeation process
for different $\lambda$-values.

As the lateral correlation length $\xi$ increases close to the critical temperature, the life-time of lipid
clusters also increases due to critical slowing down, and the typical relaxation time scales with $\xi$ with 
a dynamical critical exponent~\cite{Smith12}.
The increased life-time of lipid clusters at criticality certainly enhances the permeation process, and 
shall be further investigated in future studies.
}

We remark that the obtained temperature dependency of the permeation rate should
be distinguished from the previously discussed Arrhenius-type behavior of the skin
permeability coefficient~\cite{Scheuplein1971}.
In experiments, the permeability coefficient was shown to obey the form
$K_{\rm p} \sim \exp(-E_{\rm a}/k_{\rm B} T)$ where $E_{\rm a}$ is an activation energy
of solute molecules or ions.
Hence, $K_{\rm p}$ increases with temperature, and manifests an opposite trend when compared
with our result, eq.~(\ref{v_eT}).
In our work we have focused on the enhanced permeability due to the concentration fluctuations,
which are very sensitive to the proximity to the critical temperature $T_{\rm c}$.
If lateral heterogeneities in living biomembranes at physiological conditions correspond to
critical fluctuations~\cite{Smith08,Veatch08,Smith12}, we expect that the material
transport through clusters should also be significantly affected around physiological
temperatures, as studied in this Letter.

A typical value of the permeability coefficient through the stratum corneum was
measured~\cite{Potts92} to be $K_{\rm p} \approx 5 \times 10^{-8}$\,m/s.
Assuming that the inter-membrane distance is about $d \approx 5 \times 10^{-9}$\,m, we
can estimate a characteristic time scale for the permeation as $d/K_{\rm p} \approx 0.1$\,s,
which is fairly large.
Although we cannot yet make a direct connection with the experiments, the waiting time $t_w$
in our model should be
comparable to this time scale.
From a microscopic point of view, the waiting time $t_w$ is determined by the hydrophobic
interactions between the permeant molecule and the surrounding membrane environment.

Finally, we note that the self-diffusion coefficient of ions through a \textit{single-component
unilamellar} vesicle was shown to increase substantially near the lipid \textit{main-transition}
temperature~\cite{Papahadjopoulos}.
In those studies, the phase transition occurs between two states of lipid molecules:
an ordered gel phase (solid-like phase) and a disordered liquid crystalline phase (liquid-like phase).
Furthermore, some models suggested~\cite{Cruzeiro-Hansson,J. F. Nagle,Yang2015}
that the interfaces between the gel and disordered-liquid domains of the lipid molecules are
responsible for the high molecular permeability.
Our model differs from these models, because we have focused on the enhanced permeability due
to concentration fluctuations in a \textit{multi-component lamellar stack} close to the critical
temperature of the binary lipid mixture.

\section{Conclusions}

To summarize, we have performed Monte Carlo simulations of a stack of binary lipid mixture
arranged in a multi-layered structure, and considered the transport of permeant molecules
through such a lamellar stack.
Within our model, permeants can be transported to adjacent membranes through
vertically connected lipid clusters that are transiently formed  by lipid concentration fluctuations
in the one-phase above $T_{\rm c}$.
We have found that the permeation rate decays exponentially with temperature and with the
permeant cross-sectional area, whereas the dependency on the waiting time obeys a stretched
exponential behavior.
Such an exponential dependence on  temperature is analogous to the behavior of the in-plane
correlation function, and in accordance with the experimentally observed dependency of the
permeability coefficient on permeant size.
Our results imply that concentration fluctuations in physiological conditions can play an
important role for efficient material transport through multi-component membranes.

It may be of interest to include in future works the excluded volume effect of the permeant molecules,
as well as the interaction between permeants, in order to broaden the scope of the present model
and connect it more directly to permeation processes through multi-layered biological membranes.

\acknowledgments

We thank T.\ Kato, R.\ Okamoto and T.\ V.\ Sachin Krishnan for helpful discussions.
T.H.\ acknowledges support by Grant-in-Aid for JSPS Fellows　
(Grant No.\ 17J01643)　from the Japan Society for the Promotion of Science (JSPS).
S.K.\ acknowledges support by Grant-in-Aid for Scientific Research on Innovative 
Areas ``\textit{Fluctuation and Structure}" (Grant No.\ 25103010) from the 
Ministry of Education, Culture, Sports, Science, and Technology (MEXT) of Japan,
and by Grant-in-Aid for Scientific Research (C) (Grant No.\ 15K05250) from the JSPS.
D.A.\ acknowledges support from the Israel Science Foundation (ISF) under grant No.\ 438/12,
the U.S.--Israel Binational Science Foundation (BSF) under grant No.\ 2012/060, and the ISF-NSFC
joint research program under grant No.\ 885/15.
He also thanks FU, Berlin, for its hospitality and the Alexander von Humboldt Foundation for a
Humboldt research award.


\end{document}